# Structural optimization of biohydrogen production: Impact of pretreatments on volatile fatty acids and biogas parameters

Mahmood Mahmoodi-Eshkaftaki[a*], Gustavo Mockaitis[b]

[a] Department of Mechanical Engineering of Biosystems, Jahrom University, P.O. Box 74135-111, Jahrom, Iran.

[b] Interdisciplinary Research Group on Biotechnology Applied to the Agriculture and the Environment, School of Agricultural Engineering, University of Campinas (GBMA/FEAGRI/UNICAMP), 501 Candido Rondon Avenue, CEP, 13.083-875, Campinas, SP, Brazil.

*** Email:** m.mahmoodi5@gmail.com, m.mahmoodi5@jahromu.ac.ir

## Abstract

The present study aims to describe an innovative approach that enables the system to achieve high yielding for biohydrogen (bio-$H_2$) production using xylose as a by-product of lignocellulosic biomass processing. A hybrid optimization technique, structural modelling, desirability analysis, and genetic algorithm could determine the optimum input factors to maximize useful biogas parameters, especially bio-$H_2$ and $CH_4$. As found, the input factors (pretreatment, digestion time and biogas relative pressure) and volatile fatty acids (acetic acid, propionic acid and butyric acid) had significantly impacted the biogas parameters and desirability score. The pretreatment factor had the most directly effect on bio-$H_2$ and $CH_4$ production among the factors, and the digestion time had the most indirectly effect. The optimization method showed that the best pretreatment was acidic pretreatment, digestion time > 20 h, biogas relative pressure in a range of 300–800 mbar, acetic acid in a range of 90–200 mg/L, propionic acid in a range of 20–150 mg/L, and butyric acid in a range of 250–420 mg/L. These values caused to produce $H_2$ > 10.2 mmol/L, $CH_4$ > 3.9 mmol/L, $N_2$ < 15.3 mmol/L, $CO_2$ < 19.5 mmol/L, total biogas > 0.31 L, produced biogas > 0.10 L, and accumulated biogas > 0.41 L.

## Introduction

The environmental compliance concerning reducing greenhouse gases has attracted the interest in non-conventional fuel from bioresources and wastes. Anaerobic digestion converts the wastes into valuable sources, produces biogas as a clean energy source, and reduces the waste volumes in parallel [1]. Agricultural by-products, the most important wastes in the environment, comprise cellulose, hemicellulose and lignin. Xylose units derive from hemicellulose degradation and can be used to produce second-generation biofuels because of their worldwide abundance and carbon neutrality [2].

Bio-$H_2$ is a clean fuel that does not release carbon dioxide during combustion, and it is applicable in fuel cells for electricity generation [3]. Moreover, $H_2$ has the highest energetic content (122 kJ/g) when compared to hydrocarbon fuels. Therefore, biological hydrogen production technologies, including photo and dark fermentation, direct and indirect photolysis, have received considerable attention in recent years. Bio-$H_2$ production via dark fermentation has become a promising technology due to the versatility of feedstocks that can be used, high production rates, and simple bioreactor technologies used in the production [4]. Volatile fatty acids (VFAs) obtained as by-products from the digestion can be transformed into liquid biofuels such as ethanol and butanol by controlling the digestion or transformed into biogas compounds such as $CH_4$, $H_2$, $CO_2$, and $H_2S$ [5]. Therefore, optimum amounts of the VFAs in the anaerobic digestion are essential to control the digestion. Even though dark fermentation is a high-potential technology to process waste into bio-$H_2$ [2], it needs to be improved with the optimum fermentation condition.

Different parameters influence bio-$H_2$ and $CH_4$ production like digestion time, biogas pressure, substrate degradability, feedstock pH, and VFAs contents. To improve the substrate degradability and feedstock pH, physical, chemical, and biological pretreatments are used [6]. The pretreatments render the substrates fit for greater biogas yields. The difficulty of bioprocessing lignocellulosic biomasses is related to the intrinsic characteristics of the biomass sources. The factors contributing to biomass resistance to biodegradation include crystallinity and degree of polymerization of cellulose, porosity, additional cellulose protection by lignin, sheathing of cellulose by hemicelluloses and fiber strength [7, 8]. Pretreatments break down biomass structures, and the hydrolysis stage of the anaerobic digestion is improved. However, the effect of pretreatments on bio-$H_2$ production has been studied previously [2, 3, 9, 10], optimization of the pretreatments is very important.

To determine the best method to control the effective parameters in bio-$H_2$ production, optimization methods based on regression models like response surface methodology (RSM) show good accuracy [3, 9, 11]. However, the RSM alone is not very accurate in optimizing multiple responses, especially when some moderate factors such as VFAs influence the response factors. Intelligent methods like integrated artificial neural network and genetic algorithm (GA) show high accuracy for such problems [12]. The novelty of this article is to develop a hybrid optimization method, integrated regression models and GA, to determine the best condition of bio-$H_2$ production by overlaying different responses. To that end, researchers employed a structural equation model (SEM) to quantify the relative contribution of the process parameters to the biogas production desirability, developed multivariate regression models to estimate the biogas parameters, and integrated the desirability analysis (DA) and the GA to determine the optimum amounts.

## Materials and methods

### *Inoculum pretreatments*

The inoculum used in this study was an anaerobic granular sludge obtained from a UASB reactor treating poultry slaughterhouse wastewater. The concentration of total volatile solids (T.VS) in the sludge was 32 g/L, with a pH of 7.1. In the experiments, the sludge granules were macerated to improve the external mass transfer, increase the superficial area of the bioparticles, and make the pretreatments more effective. Four different pretreatments were performed on the inoculum: acidic, thermal, acidic-thermal, and thermal-acidic compared to a control experiment, which were used with the numbers 1, 2, 3, and 4 in comparison with 0 in the modelling and optimization processes. Acidic pretreatment was performed by adding HCl solution (1 mol/L) until the pH of the sludge dropped down to 3.0 and maintained under these conditions for 24 h. Afterwards, the pH was raised to 6.0 with the addition of NaOH solution (1 mol/L). The thermal pretreatment was done by raising the temperature of the sludge up to 90 °C for 20 min, and then the sludge was quickly cooled down to 25 °C with a water bath. Acidic-thermal and thermal-acidic pretreatments were a combination of acid and thermal pretreatments described above, which were conducted as indicated in sequence, after 24 h at room temperature between the two pretreatments [2, 13, 14]. The control experiments were done without performing any pretreatment. For all pretreated and control sludges, the pH was set to 6.5 before their inoculation in the experiments. The pretreated sludge was kept at room temperature for 24 h before further inoculation.

***Medium***

Xylose was the sole carbon source used in all assays (7.5 ± 0.6 g/L). Nutrient supplementation was prepared as described in Mockaitis et al. [2], and contained $NiSO_4$ (0.5 mg/L), $FeSO_4$ (2.5 mg/L), $FeCl_3$ (0.25 mg/L), $CoCl_2$ (0.04 mg/L), $CaCl_2$ (2.06 mg/L), $SeO_2$ (0.04 mg/L), $KH_2PO_4$ (1.30 mg/L), $KHPO_4$ (5.36 mg/L), and $Na_2HPO_4$ (2.76 mg/L). Further, urea was used as a nitrogen source (62 mg/L). pH of the medium was initially set to 6.5 using an HCl solution.

***Experimental setup***

All experiments were performed in 500 mL Duran™ Flasks (total volume of 620 ± 14 mL), with a working volume of 285 ± 8 mL (medium + inoculum), and a headspace volume of 335 ± 18 mL. The headspaces of the reactors were purged for 3 min with $N_2$ 99% and sealed immediately [15]. In all assays, the initial inoculum concentration was 6.6 ± 0.3 g $\mathrm{T.VS/L}$, and the initial pH was 6.68. The experiments were carried out under mesophilic conditions (30 °C) in a shaker incubator with orbital stirring at 150 rpm.

***Dataset***

For each pretreatment, the biogas compounds ($H_2$, $CO_2$, $N_2$, $CH_4$) were measured throughout gas chromatography using a GC Shimadzu™ GC2010, according to the procedure established by Peixoto et al. [16]. The concentrations of VFAs (acetic, propionic, butyric acids) were determined using a gas chromatograph (GC) Shimadzu™ GC2010 with a flame ionization detector and an HP-INNOWAX column of 30 m × 0.25 mm (internal diameter) × 0.25 mm (film thickness). Some biogas yield parameters including total biogas, produced biogas, and accumulated biogas were also measured. The data for each pretreatment were collected over time for 169 h. The trend of bio-$H_2$ production versus digestion time and relative pressure of biogas production was considered for each pretreatment, and 28 relative maximums were extracted for each pretreatment. Fig. 1 shows the extracted points of relative maximums for the acidic pretreatment. For these points, the digestion time, biogas relative pressure, VFAs, biogas compounds, and biogas yield parameters were achieved. These points were extracted for the other pretreatments. Totally, a dataset with size of 140×13 was determined, in which the input factors were digestion time, biogas relative pressure and pretreatments, the moderate factors were acetic, propionic and butyric acids, and the responses were biogas compounds and biogas yield parameters. The summary of the experimental data is reported in Table 1.

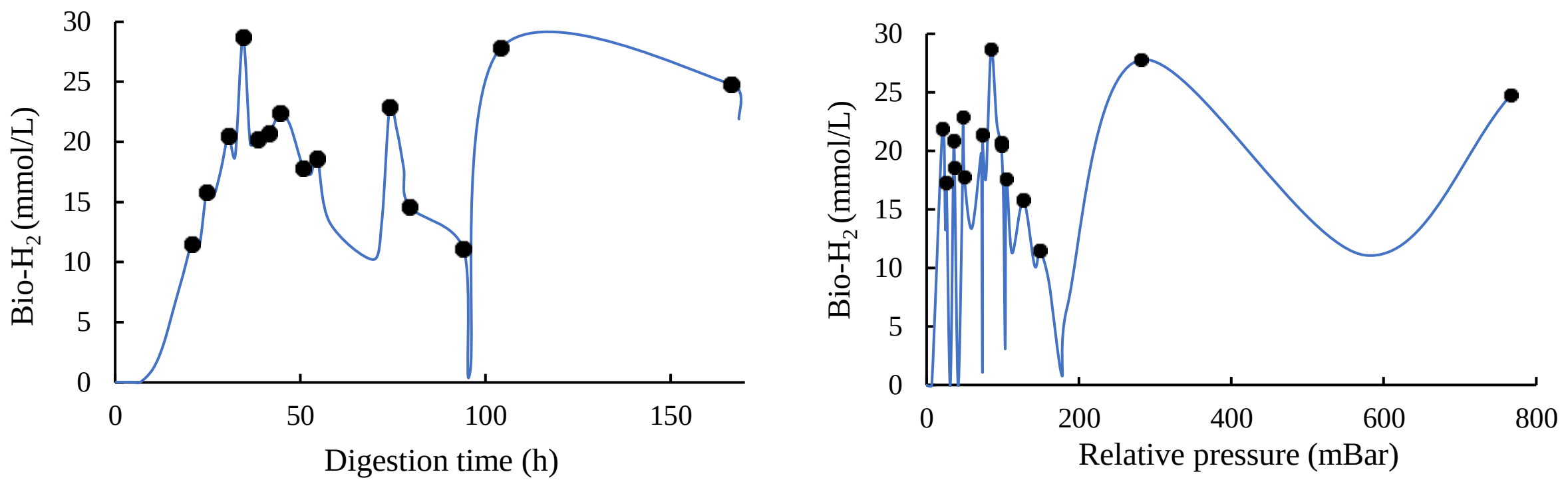

Figure 1 – Relative maximums of bio-$H_2$ achieved for the acidic pretreatment used as the input dataset in the optimization process.

Table 1 – Statistical information of the experiments including independent variables, moderate variables and dependent variables.

| | Factors | Parameters | Range | Mean | Std. Dev. |
|---|---|---|---|---|---|
| **Independent variables** | $x_1$ | Pretreatments | 0–4 | 2 | 1.414 |
| | $x_2$ | Digestion time (h) | 0.6–168.2 | 56.096 | 40.381 |
| | $x_3$ | Relative pressure (mbar) | 0–1076.6 | 154.867 | 222.119 |
| **Moderate variables** | $x_4$ | Acetic acid (mg/L) | 3–1672 | 551.993 | 401.259 |
| | $x_5$ | Propionic acid (mg/L) | 0–282 | 25.225 | 55.179 |
| | $x_6$ | Butyric acid (mg/L) | 3–1984 | 806.249 | 562.714 |
| **Dependent variables** | $y_1$ | $H_2$ (mmol/L) | 0–36.9 | 10.581 | 9.243 |
| | $y_2$ | $CH_4$ (mmol/L) | 0–47.5 | 17.224 | 12.174 |
| | $y_3$ | $N_2$ (mmol/L) | 0–17.1 | 1.686 | 3.936 |
| | $y_4$ | $CO_2$ (mmol/L) | 0–35.8 | 17.349 | 9.266 |
| | $y_5$ | Total biogas (L) | 0.254–0.583 | 0.313 | 0.066 |
| | $y_6$ | Produced biogas (L) | 0–0.314 | 0.045 | 0.064 |
| | $y_7$ | Accumulated biogas (L) | 0.008–1.264 | 0.387 | 0.302 |

***Optimization process***

This study introduces an efficient optimization method by combining regression modelling, DA, and GA to find the optimum conditions of anaerobic digestion to produce high amounts of bio-$H_2$, $CH_4$, total biogas, produced biogas and accumulated biogas, and low amounts of $CO_2$ and $N_2$. The desirability function was developed by overlaying the regression models with different relative importance. It was used as a fitness function of GA to determine the optimum amounts. The desirability function was used to simultaneously optimize all affecting parameters to achieve the best conditions of digestion. This approach is widely used, especially in the industry, to optimize multiple response processes [17]. The method finds operating conditions $x$ providing the most desirable response values [17]. For each response of $y_i(x)$, the desirability function $d_i(y_i)$ got values between 0 to 1 to the possible values of $y_i$; $d_i(y_i) = 0$ represented an undesirable value of $y_i$, and $d_i(y_i) = 1$ representing a highly desirable value.

These values were calculated using the method described by Mahmoodi-Eshkaftaki and Ebrahimi [12]. To calculate the desirability values, predicted response values ($\hat{y}_i$) was used instead of the $y_i$ [18]. A quadratic model (Eq. 1) is generalized to all the response variables. The overall desirability was calculated using the geometric mean of the individual desirability levels (Eq. 2).

$$y(x) = a_0 + \sum_{i=0}^{n} a_i x_i + \sum_{i=0}^{n} a_{ii} x_i^2 + \sum_{i<j}^{n} a_{ij} x_i x_j + ....x_i (i = 1, 2, ....n) \quad (1)$$

$$D = \left[ \prod_{i=1}^{n} d_i (\hat{y}_i)^{r_i} \right]^{\frac{1}{\sum r_i}} \quad (2)$$

In which $n$ is the number of responses (7 in this study), and $r_i$ is the relative importance of a response among the responses. The $r_i$ values of bio-$H_2$, $CH_4$, $N_2$, $CO_2$, total biogas, produced biogas, and accumulated biogas were 5, 4, 1, 3, 2, 4, and 1, respectively, indicating their relative importance for a high performance of the anaerobic digestion. Bio-$H_2$, $CH_4$, total biogas, produced biogas, and accumulated biogas should be maximized, and the other parameters should be minimized. Bio-$H_2$ concentration had the highest importance ($r_i = 5$) and after that were $CH_4$ concentration and produced biogas ($r_i = 4$).

To determine optimal values of a specific problem using the GA, a candidate solution is called an individual or a chromosome. Each individual represents a point in the search space, and hence a possible solution to the problem. Each individual is decided by an evaluation mechanism to obtain its fitness value. Based on this fitness value and undergoing genetic operators, a new population is generated iteratively, with each successive population referred to as a generation [20]. In this study, the roulette wheel method was used for chromosome selection. The crossover operator was applied to create a pair of offspring chromosomes. A two-point crossover operation was used for each selected pair to generate an offspring and a one-gene mutation operation was applied to generate new chromosomes. Different crossover and mutation probabilities were considered using MATLAB (Ver. 7.12) software to improve the GA performance.

***Statistical analysis***

Six input factors (independent variables and moderate variables) and seven responses were acquired from the experiments. The effect of independent variables on the moderate variables and their effect on the responses were statistically calculated with one-way ANOVA using SPSS (Ver. 22). The SEM was developed to evaluate the hypothetical responses of biogas compounds and biogas yield parameters to the input factors and determine the effect of the biogas compounds and biogas yield parameters on the desirability factor. SEM is a priori approach with the capacity to identify causal relationship between variables by fitting data to the models representing causal hypotheses. The obtained correlation matrix used for model fitting was implied in AMOS (Ver. 24) software to construct the SEM using the maximum-likelihood estimation method. Non-significant chi-square test ($P > 0.05$), high goodness-of-fit index ($GFI > 0.53$) and low root mean square errors of approximation ($RMSEA < 0.35$) indicated the goodness of fit of the SEM. The standardized total effects of all the variables were also calculated among the variables during the DA.

## Results and discussion

As shown in Table 1, the responses produced relatively wide ranges of the biogas parameters that were suitable for model development and optimization process. The desirability function (Eq. 2) should be maximized (or $\frac{1}{\text{Eq.2}}$ should be minimized) to optimize these parameters. The most crucial goal of this study was optimizing the bio-$H_2$ production; therefore, the response of bio-$H_2$ had the most relative importance in the DA. The biogas parameters were modelled according to the input factors, and the models were overlayed in the desirability function.

***Synergistic mechanisms of input factors – desirability***

The SEM was used to quantify the relative contribution of the process parameters to the changes of desirability factor (Fig. 2). SEM analysis appeared that the independent variables together could explain 93% of the change of VFAs, and the independent variables and VFAs together could explain 99% of the change of biogas yield parameters and 93% of the change of biogas compounds. The biogas yield parameters and the biogas compounds together explained 61% of the change of desirability factor. As shown in Fig. 2, the digestion time and biogas relative pressure indirectly impacted biogas compounds by significantly affecting VFAs ($\lambda = -1.06$, P < 0.001, and $\lambda = 0.16$, P< 0.01, respectively). The pretreatment factor directly and significantly impacted the biogas compounds ($\lambda = 0.28$, P < 0.001) and the biogas yield parameters ($\lambda = 0.01$, P < 0.05). VFAs ($\lambda = 1.76$, P < 0.01) directly impacted biogas compounds and indirectly impacted the desirability factor. The biogas compounds and biogas yield parameters directly impacted the desirability factor ($\lambda = -0.6$, P < 0.001, and $\lambda = -0.45$, P< 0.001, respectively). These results indicated that the input factors indirectly and significantly impacted the desirability factor. On the other hand, bio-$H_2$ had the highest relative importance to calculate the desirability value, confirming that the input factors had a significant effect on bio-$H_2$ production. The SEM can be used to estimate the strength of these multiple (direct and indirect) effects, including the standardized direct, indirect and total effects (Fig. 2). The digestion time had a negative effect on VFAs, while the biogas relative pressure had positive effects. Different trends between VFAs and digestion time have been reported in the literature [21, 22]. The pretreatment had a positive effect on VFAs, biogas compounds, and biogas yield parameters. The results suggested that the thermal-acidic pretreatment (marked with the number 4) had the best improvement in the desirability of biogas production and after that was acidic-thermal pretreatment (marked with the number 3). As shown in Fig. 2, to increase the desirability factor and optimize all the responses, the most effective VFAs were propionic and butyric acids, the most effective biogas compounds were bio-$H_2$ and $CO_2$ compounds, and the most effective biogas yield parameter was produced biogas.

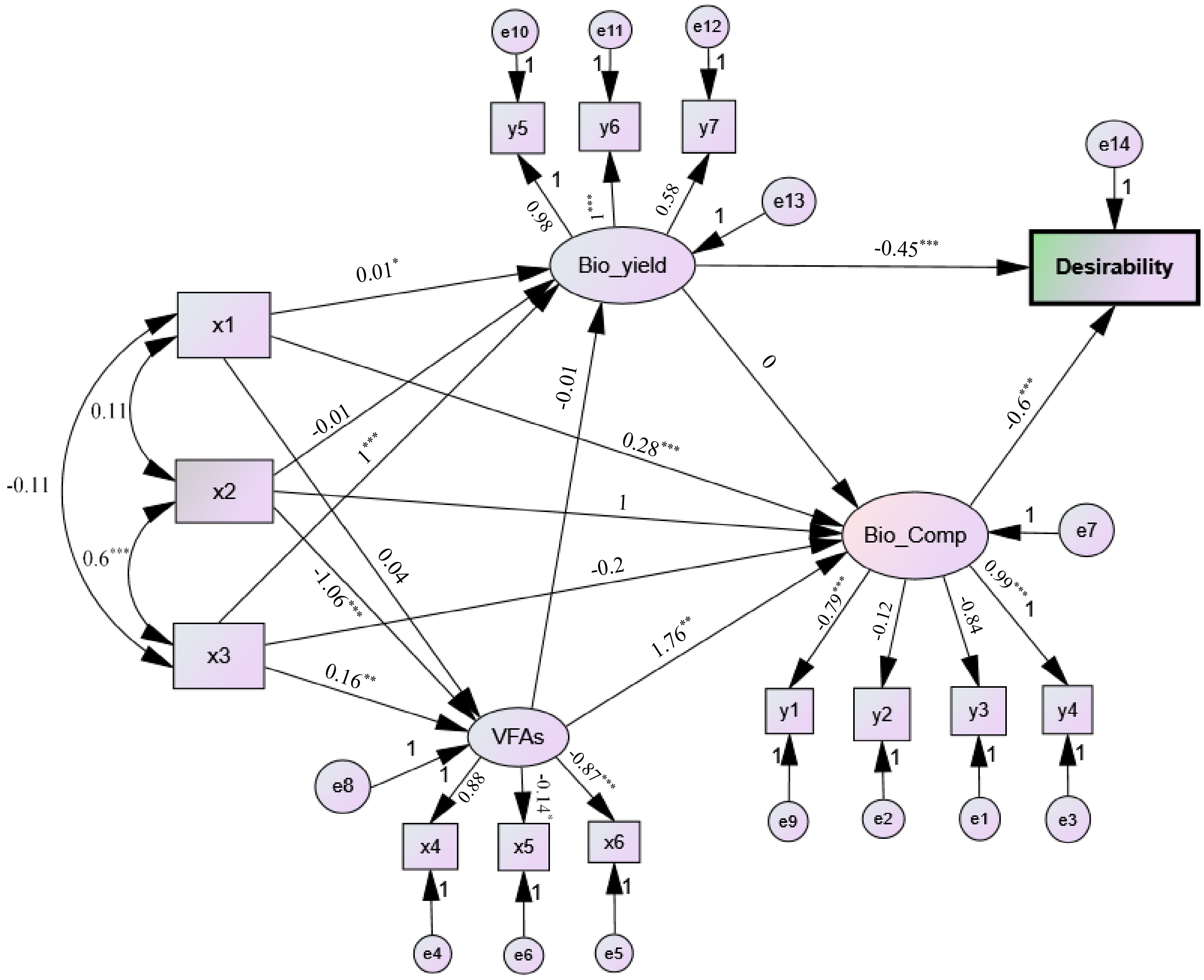

Figure 2 – Structural equation model showing the direct and indirect effects of independent variables ($x_1$: pretreatments, $x_2$: digestion time, $x_3$: relative pressure), moderate variables ($x_4$: acetic acid, $x_5$: propionic acid, $x_6$: butyric acid), dependent variables ($y_1$: $H_2$, $y_2$: $CH_4$, $y_3$: $N_2$, $y_4$: $CO_2$, $y_5$: total biogas, $y_6$: produced biogas, $y_7$: accumulated biogas), and desirability factor. The numbers adjacent to the arrows are standardized path coefficients and indicative of the effect size of the relationship. $^{*}P < 0.05$, $^{**}P < 0.01$ and $^{***}P < 0.001$.

The initial amount of VFAs was higher for all the pretreatments, as compared to the control, suggesting that the pretreatments generated initial metabolic stress on the original inoculum. The control experiment was the only assay showing a significant production of propionate, which could indicate a possible instability of the anaerobic process. The acidic pretreatment showed the highest maximum production rate for $H_2$ production. This pretreatment showed a higher biological activity as compared to all other assays, including control, except for the methanogenic activity, which was suppressed. Except for the thermal one, all pretreatments showed a higher yield to convert xylose into $H_2$ than in the control experiment. The pretreatments performed in inoculum are intended to hinder methanogenesis. Thus, the strong relationship between the pretreatments performed and the low $CH_4$ amount in biogas showed that this strategy was successfully accomplished. The same relationship indicates that propionic acid production was affected by the pretreatment.

Mean comparison using one way ANOVA showed that (i) all the input factors had significant effects at level 0.01 on bio-$H_2$, $CO_2$, and $N_2$ production, (ii) the pretreatment factor and propionic acid showed significant effects at level 0.01 on $CH_4$ production, (iii) the digestion time and VFAs had significant effects at level 0.01 on the total and produced biogas, and (iv) the digestion time, biogas relative pressure and VFAs had significant effects at level 0.01 on accumulated biogas. Pearson's correlation coefficients (indicates linear correlations) showed that among the independent variables just the biogas relative pressure had significant correlation with the desirability factor ($r=-0.168^{*}$), indicating a linear correlation between the biogas relative pressure and desirability factor. The other independent variables had a positive non-significant correlation with the desirability factor. Spearman's correlation coefficients (indicates non-linear correlations) showed a significant correlation between the digestion time and desirability factor ($r=0.245^{**}$), indicating a non-linear correlation between them. The other independent variables had a positive non-significant correlation with the desirability factor. These results agree with other studies. As reported in the literature for anaerobic digesters [23, 24, 25], there are complicated and non-linear relationships among the system variables and keep them under control is complex. For example, the relationship between biogas relative pressure and VFAs might be explained due to the increase of available $CO_2$ within the liquid phase. Due to its high solubility, the rise in $CO_2$ concentration probably directed the group metabolism towards autotrophic pathways, impacting the VFAs production directly and the biogas composition indirectly [2].

***Regression modelling***

Eq. 1 was generalized for the responses according to the input factors. Coefficients of the model's terms developed for the responses are illustrated in Table 2. As found, accuracy of the model of produced biogas was the most among the models ($R^2$= 0.999, $AR^2$=0.998, MSE=0.00001). The quantities of $R^2$ and adjusted $R^2$ of the bio-$H_2$ model as the most important response were high (0.945 and 0.813, respectively), and the MSE was low (16.37). The statistical quantities calculated for the models showed high accuracy for the models to be used in the optimization process.

The coefficients were determined using the standardized dataset; therefore, the values of the coefficients illustrate their effects on each response [19, 26]. It was found from the developed models of bio-$H_2$ and $CH_4$ that the pretreatment factor had the most effect on bio-$H_2$ and $CH_4$ production among the factors, and after that were propionic acid and digestion time. As shown in the table for bio-$H_2$, the digestion time ($x_1$) linearly had no significant effect on bio-$H_2$, while non-linearly (${x_1}^2$) was significant. It was agreed with SEM results, as the digestion time had no direct impact on biogas compounds. In the developed $CH_4$ model, the digestion time was not linearly and non-linearly significant, while the pretreatment factor was linearly and non-linearly significant. This result is findable from SEM in Fig. 2. In the developed models for the other responses, the pretreatment factor showed the highest coefficients and thus had the most effect and after that was the digestion time. As shown in the table for the developed models of the biogas yield parameters, the coefficients of the most squared terms were zero indicating that the input factors linearly impacted the biogas yield parameters.

Table 2 – Model coefficients developed for the responses and accuracy of each model.

| **Factor** | **Bio-H₂** | **CH₄** | **N₂** | **CO₂** | **Total biogas** | **Produced biogas** | **Accumulated biogas** |
|---|---|---|---|---|---|---|---|
| $a_0$ | $8.95979^{**}$ (3.4762) | $1.62592^{ns}$ (1.5727) | $28.84178^{***}$ (3.9979) | $-7.51578^{*}$ (3.2461) | $0.24655^{***}$ (0.0069) | $-0.00158^{ns}$ (0.0022) | $0.18444^{**}$ (0.0670) |
| $x_1$ | $2.10540^{ns}$ (1.7004) | $-6.66578^{***}$ (0.7693) | $7.11890^{**}$ (1.9556) | $-11.59884^{***}$ (1.5878) | $0.02876^{***}$ (0.0034) | $0.00436^{***}$ (0.0011) | $-0.01468^{ns}$ (0.0328) |
| $x_2$ | $0.17085^{*}$ (0.0945) | $0.01327^{ns}$ (0.0428) | $-0.47495^{***}$ (0.1087) | $0.26050^{**}$ (0.0883) | $0.00051^{**}$ (0.0002) | $0.00011^{*}$ (0.0001) | $0.01215^{***}$ (0.0018) |
| $x_3$ | $0.00982^{*}$ (0.0056) | $0.01046^{***}$ (0.0025) | $-0.01243^{*}$ (0.0065) | $0.01346^{**}$ (0.0052) | $0.00030^{***}$ (-) | $0.00029^{***}$ (-) | $0.00019^{*}$ (0.0001) |
| $x_4$ | $-0.02323^{***}$ (0.0058) | $0.00740^{**}$ (0.0026) | $0.01381^{*}$ (0.0066) | $0.01840^{***}$ (0.0054) | $0.00002^{*}$ (-) | $-0.000002^{ns}$ (-) | $-0.00058^{***}$ (0.0001) |
| $x_5$ | $-0.17673^{***}$ (0.0218) | $0.07521^{***}$ (0.0098) | $0.08476^{**}$ (0.025) | $0.02922^{ns}$ (0.0203) | $-0.00018^{***}$ (-) | $-0.00003^{*}$ (-) | $-0.00097^{*}$ (0.0004) |
| $x_6$ | $0.01620^{**}$ (0.0058) | $0.00099^{ns}$ (0.0026) | $-0.01714^{**}$ (0.0066) | $0.03025^{***}$ (0.0054) | $-0.00003^{**}$ (-) | $-0.00001^{ns}$ (-) | $0.00013^{ns}$ (0.0001) |
| ${x_1}^2$ | $-1.06424^{**}$ (0.332) | $1.30976^{***}$ (0.1502) | $-0.70716^{*}$ (0.3819) | $2.08840^{***}$ (0.3101) | $-0.00708^{***}$ (0.0007) | $-0.00103^{***}$ (0.0002) | $-0.01727^{**}$ (0.0064) |
| ${x_2}^2$ | $-0.00089^{*}$ (0.0004) | $-0.00004^{ns}$ (0.0002) | $0.00177^{***}$ (0.0004) | $-0.00067^{*}$ (0.0004) | $-0.000002^{*}$ (-) | - (-) | $-0.00004^{***}$ (-) |
| ${x_3}^2$ | $0.00001^{*}$ (-) | $-0.00001^{**}$ (-) | $0.00001^{ns}$ (-) | $-0.000001^{ns}$ (-) | - (-) | - (-) | - (-) |
| ${x_4}^2$ | $0.00001^{***}$ (-) | $-0.000004^{**}$ (-) | $-0.00001^{**}$ (-) | $-0.000003^{ns}$ (-) | - (-) | - (-) | - (-) |
| ${x_5}^2$ | $0.00055^{***}$ (0.0001) | $-0.00035^{***}$ (-) | $-0.00016^{ns}$ (0.0001) | $-0.00025^{**}$ (0.0001) | $0.000001^{***}$ (-) | - (-) | $0.000003^{ns}$ (-) |
| ${x_6}^2$ | $-0.00001^{***}$ (-) | - (-) | $0.00001^{**}$ (-) | $-0.00001^{***}$ (-) | - (-) | - (-) | $-0.0000002^{***}$ (-) |
| $R^2$ | 0.945 | 0.850 | 0.971 | 0.982 | 0.999 | 0.999 | 0.989 |
| $AR^2$ | 0.813 | 0.791 | 0.856 | 0.840 | 0.985 | 0.998 | 0.934 |
| MSE | 16.370 | 3.352 | 21.652 | 14.270 | 0.00006 | 0.00001 | 0.0061 |

Standard error (in parenthesis) $^{ns}$ Not significant $^{***}$ Significant at 0.001 $^{**}$ Significant at 0.01 $^{*}$ Significant at 0.05

Among the 13 terms (including $a_0$, $x_i$, and $x_i^2$) existed for each model, some determined coefficients were not significant (p-value> 0.05) and better to be deleted from the model. If there are many non-significant terms in a model, the model reduction may improve the model [24]. Therefore, the terms related to the non-significant coefficients were eliminated from the models. The accuracy of the improved models is illustrated in Fig. 3. As shown, the accuracy of the improved models is high; therefore, they were overlayed in the desirability function to determine the optimum condition.

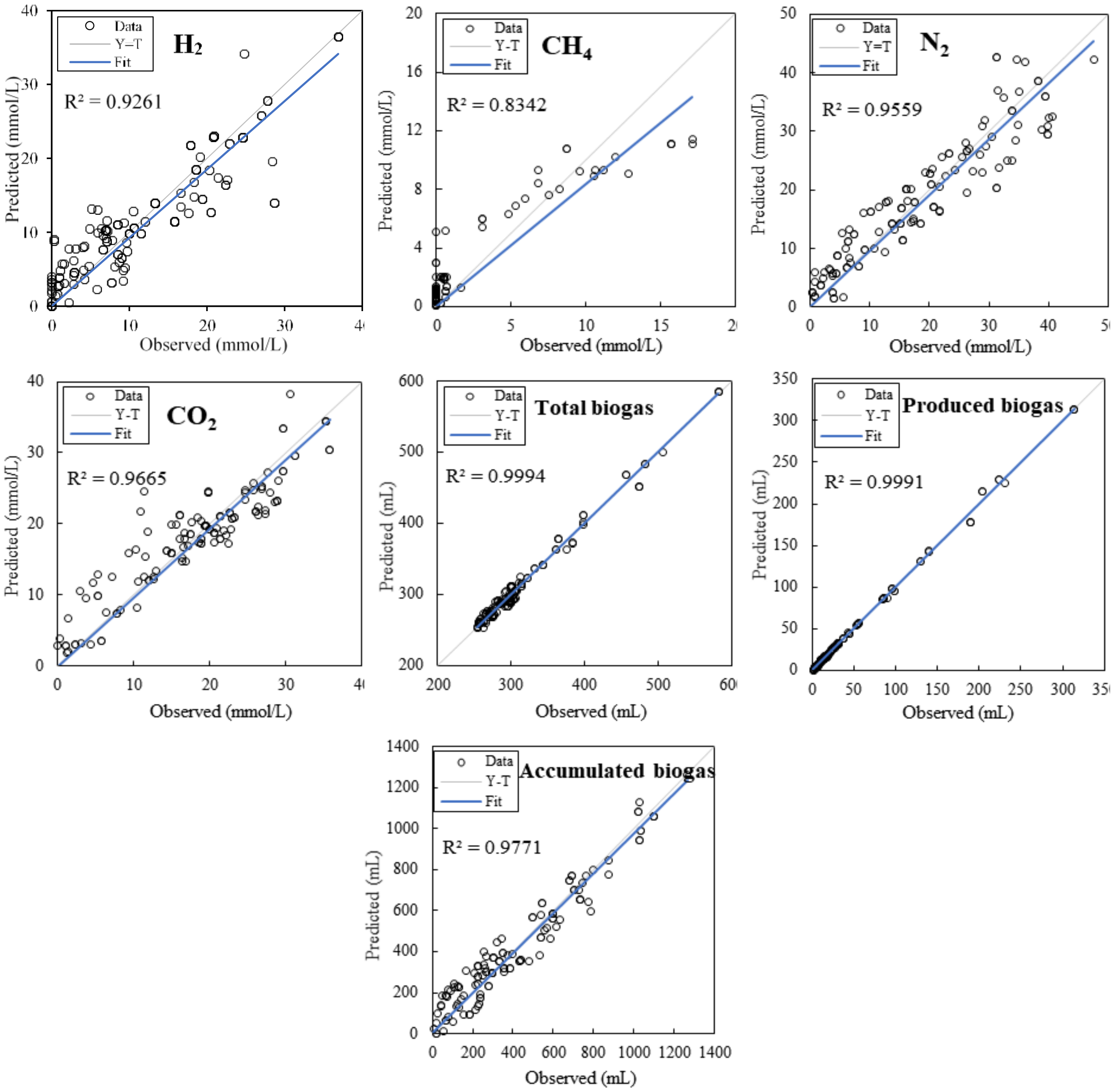

Figure 3 – Predicted values of the responses using improved models versus observed values.

***Optimization of anaerobic digestion***

Optimum amounts of the input factors to improve the anaerobic digestion were determined using the hybrid optimization method. To determine a suitable GA to be used in the hybrid optimization method, different values of population (50, 100, 150), mutation rate (0.3, 0.375, 0.45), crossover probability (0.825, 0.92, 0.975), and generation (100, 200, 500) were considered. The best ones that produced the most desirability value was employed to determine optimal amounts of the input factors and responses. GA parameters during 200 generations to minimize fitness function ($\frac{1}{\text{Desirability function}}$) are illustrated in Fig. 4.

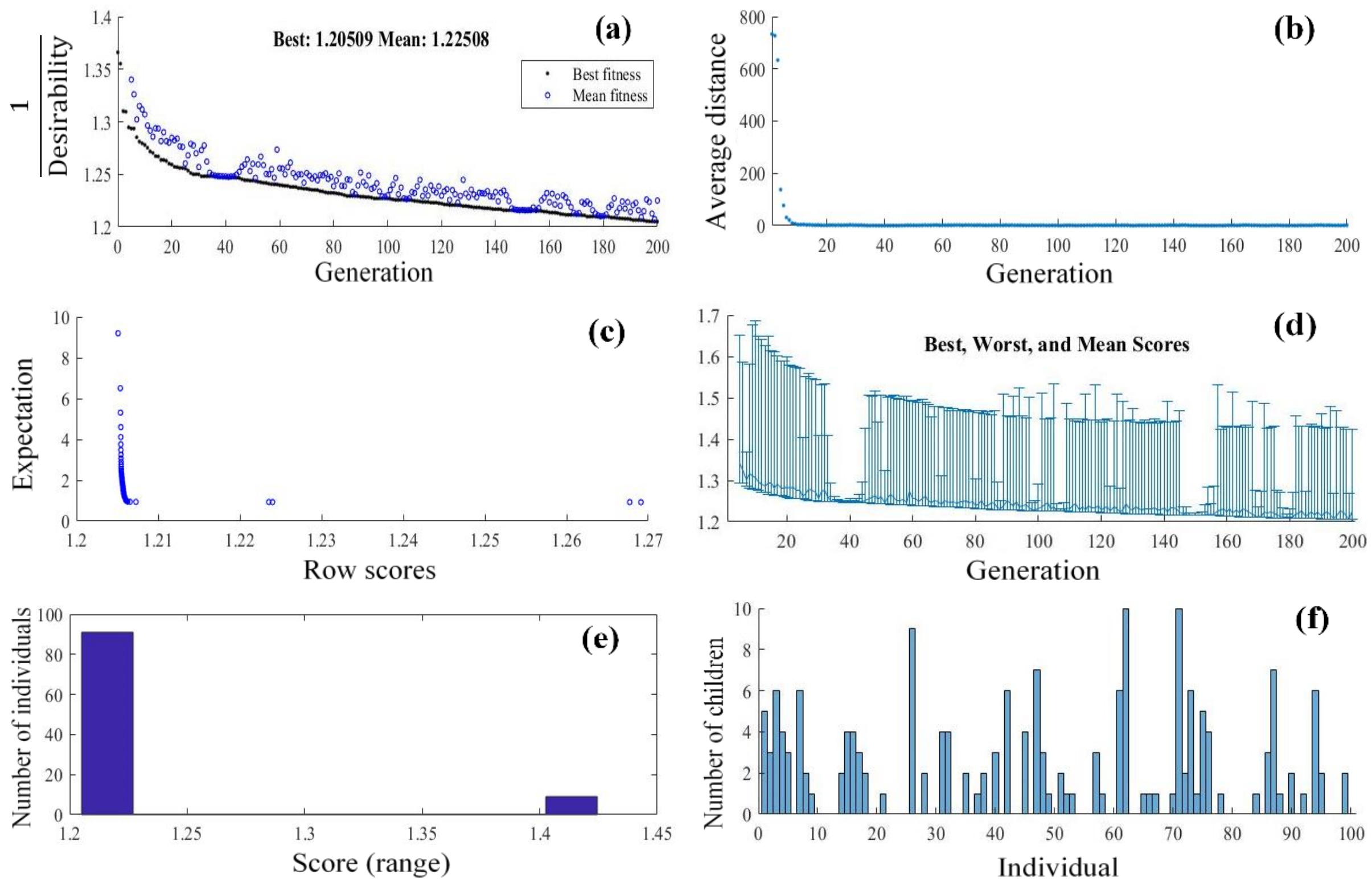

Fig. 4 – Improve GA parameters during 200 generations to optimize the desirability function; (a) Changes of the fitness function versus generation; (b) Average distance between individuals at each generation; (c) Expected number of children versus the raw scores at each generation; (d) Minimum, maximum, and mean score values in each generation; (e) Number of individuals versus scores at each generation; (f) Histogram of the parents.

As shown in Fig. 4(a), the score of $\frac{1}{\text{Desirability function}}$ decreased to the least value (1.205) up to generation 200, i.e. the desirability value increased to the most value of 0.833. A desirability value higher than 0.65 is suitable for an optimizing process [19]. It was found that with increasing the number of generations, the average distance between the individuals decreased to zero (Fig. 4b), the expected number of children decreased to the constant number of one (Fig. 4c), the changes of $\frac{1}{\text{Desirability score}}$ and its mean decreased (Fig. 4d), and the GA generated more and more similar individuals from the former generation (Fig. 4e,f). Nevertheless, several individuals continuously explored the search space, promoting a good balance between exploitation and exploration. These indicate that the GA is well developed and accurately trained.

Eighty-one runs were done by setting the GA parameters. The four best runs produced the highest desirability values were reported in Table 3. These runs were used to illustrate the effect of changes in the GA parameters and input factors on the biogas parameters. The best run (run 1) revealed the highest desirability and was selected to report the optimum amounts in this research. These optimum runs were precise and reliable because they were determined by a hybrid optimization technique, integrating accurate models, DA, and GA. Similar optimization methods have been successfully developed by integrating regression models and DA to optimize biogas production [19, 24]. However, they focused on $CH_4$ production as the major response. Optimization of anaerobic digestion using hybrid intelligent methods has been studied in recent years, an integrated ANN and GA could successfully optimize biogas production rate [24], biogas purification [12], and biogas digester shape [28]. Most of these studies focused on $CH_4$ enrichment, although in this study focused on process optimization for bio-$H_2$ production.

Table 3 – Optimal quantities of the input factors and achieved biogas parameters with four best GA parameters.

| **Factors** | Run 1 (Population: 150, Crossover: 0.92, Generation: 200, Mutation: 0.45) | Run 2 (Population: 100, Crossover: 0.825, Generation: 500, Mutation: 0.45) | Run 3 (Population: 150, Crossover: 0.92, Generation: 500, Mutation: 0.3) | Run 4 (Population: 150, Crossover: 0.825, Generation: 200, Mutation: 0.45) |
|---|---|---|---|---|
| **GA output** | | | | |
| Fitness score (Desirability value) | 1.014 (0.986) | 1.088 (0.919) | 1.089 (0.918) | 1.117 (0.895) |
| Exit flag | 0 | 0 | 0 | 0 |
| Optimization terminated | Generations exceeded | Generations exceeded | Generations exceeded | Generations exceeded |
| **Input factors** | | | | |
| Pretreatments | 1 | 0 | 0 | 1 |
| Digestion time (h) | 26.946 | 37.336 | 38.479 | 55.735 |
| Relative pressure (mbar) | 761.457 | 552.108 | 388.352 | 728.729 |
| Acetic acid (mg/L) | 155.413 | 179.805 | 101.349 | 145.798 |
| Propionic acid (mg/L) | 71.450 | 30.776 | 27.775 | 133.639 |
| Butyric acid (mg/L) | 386.007 | 285.492 | 350.282 | 283.363 |
| **Responses** | | | | |
| $H_2$ (mmol/L) | 17.517 | 17.524 | 17.229 | 14.704 |
| $CH_4$ (mmol/L) | 4.632 | 4.570 | 4.387 | 4.614 |
| $N_2$ (mmol/L) | 13.664 | 9.128 | 7.945 | 11.272 |
| $CO_2$ (mmol/L) | 15.726 | 16.453 | 16.094 | 15.019 |
| Total biogas (L) | 0.486 | 0.429 | 0.372 | 0.492 |
| Produced biogas (L) | 0.221 | 0.162 | 0.114 | 0.213 |
| Accumulated biogas (L) | 0.464 | 0.559 | 0.570 | 0.716 |

As depicted in Table 3, the highest desirability value (0.986) was achieved with population size 150, crossover probability 0.92, generation 200, and mutation 0.45. The desirability value was very high in comparison with the achieved values in other studies [12, 19, 27]. They integrated multivariate regression models and DA to determine the optimal values. In this research the GA was used in addition to the regression models and DA. This shows the GA was completely successful to improve the optimization method. The optimum amounts of input factors of pretreatments, digestion time, relative pressure, acetic, propionic, and butyric acid

were acidic pretreatment, 26.946 h, 761.457 mbar, 155.413 mg/L, 71.450 mg/L, and 386.007 mg/L, respectively. Corresponding amounts of the responses of bio-$H_2$, $CH_4$, $N_2$, $CO_2$, total biogas, produced biogas, and accumulated biogas were 17.517 mmol/L, 4.632 mmol/L, 13.664 mmol/L, 15.726 mmol/L, 0.486 L, 0.221 L, and 0.464 L, respectively. Consideration of 16 best runs of the GA showed that the maximum desirability was achieved for most runs with the acidic pretreatment, digestion time > 20 h, relative pressure in a range of 300–800 mbar, acetic acid in a range of 90–200 mg/L, propionic acid in a range of 20–150 mg/L, and butyric acid in a range of 250–420 mg/L. These values caused to produce $H_2$ > 10.2 mmol/L, $CH_4$ > 3.9 mmol/L, $N_2$ < 15.3 mmol/L, $CO_2$ < 19.5 mmol/L, total biogas > 0.31 L, produced biogas > 0.10 L, and accumulated biogas > 0.41 L. As found above, the best pretreatment was acidic pretreatment which was in agreement with the findings reported in the literature [29]. They reported that the pH < 5.5 better enhanced bio-$H_2$ production. Some optimum ranges of the biogas compounds were reported in the literature [12, 19, 24], although they were determined by focusing on $CH_4$ production as the major response. It has been found that HRT (hydraulic retention time) increasing is directly proportional with bio-$H_2$ production. However, after certain HRT, the production of bio-$H_2$ decreases [30, 31]. A similar trend was found for digestion time in this research.

It is vital to optimize multiple responses for the successful operation of a biogas plant. The RSM is useful for studying the optimum conditions of a single response only [17], while in desirable product development, several responses may have to be optimized simultaneously. Some responses must be maximized or minimized, while others may be kept within a range. The influence of each factor on a single response can be elucidated by modelling the response. Hence, individual models needed to be overlaid in the DA to study the parameters altogether. The DA can describe the complicated and non-linear relationships among the system variables. Therefore, an integrated DA–GA approach can determine the optimum conditions of multivariate non-linear systems such as biogas plants. It is practically impossible to keep all the digestion conditions at constant levels for biogas production. Even though the environment is controlled using professional equipment to a certain extent. Therefore, the optimum ranges determined in this research would considerably increase the cost of operations in this study. Develop a hybrid optimization technique, SEM–DA–GA, would be a handy tool for the operators in such situations.

**Conclusion**

The relationship among the system variables was successfully described using SEM. The SEM showed that the digestion time and biogas relative pressure indirectly impacted biogas compounds by significantly affecting VFAs, while the pretreatments directly and significantly impacted the biogas parameters. The VFAs directly impacted the biogas compounds and indirectly impacted the desirability factor, among them the propionic and butyric acids had the most effect. These results indicated that the input factors had indirectly and significantly impacted the desirability factor, and in fact they significantly affected bio-$H_2$ production. The predicted amounts of the responses using the developed models were used in the optimization process instead of the observed amounts to cover ranges outside our measured values. The hybrid optimization method, SEM–DA–GA, could determine the highest desirability (0.986) with population size 150, crossover probability 0.92, generation 200, and mutation 0.45. A very high value of desirability factor indicates that the optimization process is completely successful to determine the optimum amounts of input factors and responses. The optimum amounts of pretreatment, digestion time, biogas relative pressure, acetic acid, propionic acid, and butyric acid were acidic pretreatment, 26.946 h, 761.457 mbar, 155.413 mg/L, 71.450 mg/L, and 386.007 mg/L, respectively. Corresponding amounts of the responses of $H_2$, $CH_4$, $N_2$, $CO_2$, total biogas, produced biogas, and accumulated biogas were 17.517 mmol/L, 4.632 mmol/L, 13.664 mmol/L, 15.726 mmol/L, 0.486 L, 0.221 L, and 0.464 L, respectively. The optimum amounts of different factors based on biogas parameters were studied by focusing on bio-$H_2$ production in a laboratory scale, but there may be other effective parameters affecting the industrial plants, which should be investigated in further studies.

**References**

[1] Achinas S, Jan G, Euverink W. Theoretical analysis of biogas potential prediction from agricultural waste. Resource-Efficient Technologies 2016;2:143–7. https://doi.org/10.1016/j.reffit.2016.08.001.

[2] Mockaitis G, Bruant G, Guiot SR, Peixoto G, Foresti E, Zaiat M. Acidic and thermal pre-treatments for anaerobic digestion inoculum to improve hydrogen and volatile fatty acid production using xylose as the substrate. Renew Energy 2020;145:1388–98. https://doi.org/10.1016/j.renene.2019.06.134.

[3] Toledo-Cervantes A, Villafan-Carranza F, Arreola-Vargas J, Razo-Flores E, Oscar Mendez-Acosta H. Comparative evaluation of the mesophilic and thermophilic biohydrogen production at optimized conditions using tequila vinasses as substrate. Int J Hydrog Energy 2020;45:11000–10. https://doi.org/10.1016/j.ijhydene.2020.02.051.

[4] Lin CY, Nguyen TML, Chu CY, Leu HJ, Lay CH. Fermentative biohydrogen production and its by-products: a mini review of current technology developments. Renew Sustain Energy Rev 2018;82:4215–20. https://doi.org/10.1016/j.rser.2017.11.001.

[5] Steinbusch KJJ, Hamelers HVM, Buisman CJN. Alcohol production through volatile fatty acids reduction with hydrogen as electron donor by mixed cultures. Water Res 2008;42:4059–66. https://doi.org/10.1016/j.watres.2008.05.032.

[6] Ali Shah F, Mahmood Q, Rashid N, Pervez A, Raja IA, Maroof Shah M. Co-digestion, pretreatment and digester design for enhanced methanogenesis. Renew Sustain Energy Rev 2015;42:627–42. https://doi.org/10.1016/j.rser.2014.10.053.

[7] McKendry P. Energy production from biomass (part 1): overview of biomass. Bioresour Technol 2002;83:37–46. https://doi.org/10.1016/S0960-8524(01)00118-3.

[8] Mosier N, Wyman C, Dale B, Elander R, Lee Y, Holtzapple M, et al. Features of promising technologies for pretreatment of lignocellulosic biomass. Bioresour Technol 2005;96:673–86. https://doi.org/10.1016/j.biortech.2004.06.025.

[9] Swathy R, Rambabu K, Banat F, Ho SH, Chu DT, Show PL. Production and optimization of high grade cellulase from waste date seeds by Cellulomonas uda NCIM 2353 for biohydrogen production. Int J Hydrog Energy 2020;45:22260–70. https://doi.org/10.1016/j.ijhydene.2019.06.171.

[10] Sarkar O, Rova U, Christakopoulos P, Matsakas L. Influence of initial uncontrolled pH on acidogenic fermentation of brewery spent grains to biohydrogen and volatile fatty acids production: optimization and scale-up. Bioresour Technol 2021;319:124233. https://doi.org/10.1016/j.biortech.2020.124233.

[11] Yadav S, Singh V, Mahata C, Das D. Optimization for simultaneous enhancement of biobutanol and biohydrogen production. Int J Hydrog Energy 2021;46:3726–41. https://doi.org/10.1016/j.ijhydene.2020.10.267.

[12] Mahmoodi-Eshkaftaki M, Ebrahimi R. Integrated deep learning neural network and desirability analysis in biogas plants: a powerful tool to optimize biogas purification. Energy 2021;231:121073. https://doi.org/10.1016/j.energy.2021.121073.

[13] Mockaitis G, Bruant G, Guiot SR, Foresti E, Zaiat M. Dataset of anaerobic acidogenic digestion for hydrogen production using xylose as substrate: biogas production and

metagenomic data. Data Brief 2019;26:104466. https://doi.org/10.1016/j.dib.2019.104466.

[14] Mockaitis G. Acidic and thermal pretreatments for anaerobic digestion inoculum to improve hydrogen and volatile fatty acids production using xylose as the substrate. Mendeley Data, V4. 2021. doi: 10.17632/7knhxgvb4s.

[15] Wang X, Yang G, Feng Y, Ren G, Han X. Optimizing feeding composition and carbon/nitrogen ratios for improved methane yield during anaerobic co-digestion of dairy, chicken manure and wheat straw. Bioresour Technol 2012;120:78–83. https://doi.org/10.1016/j.biortech.2012.06.058.

[16] Peixoto G, Saavedra NK, Varesche MB, Zaiat M. Hydrogen production from soft-drink wastewater in an upflow anaerobic packed-bed reactor. Int J Hydrogen Energy 2011;36:8953–66. https://doi.org/10.1016/j.ijhydene.2011.05.014.

[17] Arun VV, Saharan N, Ramasubramanian V, Babitha Rani AM, Salin KR, Sontakke R, et al. Multi-response optimization of Artemia hatching process using split-split-plot design-based response surface methodology. Sci Rep 2017;7:1–13. https://doi.org/10.1038/srep40394,40394.

[18] Wang H, Li J, Cheng M, Zhang F, Wang X, Fan J, et al. Optimal drip fertigation management improves yield, quality, water and nitrogen use efficiency of greenhouse cucumber. Sci Hortic 2019;243:357–66. https://doi.org/10.1016/j.scienta.2018.08.050.

[19] Mahmoodi-Eshkaftaki M, Rahmanian-Koushkaki H. An optimum strategy for substrate mixture and pretreatment in biogas plants: potential application for high-pH waste management. Waste Manag 2020;113:329–41. https://doi.org/10.1016/j.wasman.2020.06.014.

[20] Gueguim Kana EB, Oloke JK, Lateef A, Adesiyan MO. Modelling and optimization of biogas production on saw dust and other co-substrates using artificial neural network and

genetic algorithm. Renew Energy 2012;46:276–81. https://doi.org/10.1016/j.renene.2012.03.027.

[21] Wang Y, Zhang Y, Wang J, Meng L. Effects of volatile fatty acid concentrations on methane yield and methanogenic bacteria. Biomass Bioenergy 2009;33:848–53. doi: 10.1016/j.biombioe.2009.01.007.

[22] Vasconcelos de Sa LR, Leal de Oliveira MA, Cammarota MC, Matos A, Ferreira-Leitao VS. Simultaneous analysis of carbohydrates and volatile fatty acids by HPLC for monitoring fermentative biohydrogen production. Int J Hydrog Energy 2011;1–10. doi: 10.1016/j.ijhydene.2011.08.056.

[23] Akbas H, Bilgen B, Turhan AM. An integrated prediction and optimization model of biogas production system at a wastewater treatment facility. Bioresour Technol 2015;196:566–76. http://dx.doi.org/10.1016/j.biortech.2015.08.017.

[24] Beltramo T, Klocke M, Hitzmann B. Prediction of the biogas production using GA and ACO input features selection method for ANN model. Inf Process Agric 2019;6:349–56. https://doi.org/10.1016/j.inpa.2019.01.002.

[25] Gao S, Bo C, Li J, Niu C, Lu X. Multi-objective optimization and dynamic control of biogas pressurized water scrubbing process. Renew Energy 2020;147:2335–44. https://doi.org/10.1016/j.renene.2019.10.022.

[26] Rezaei R, Ghofranfarid M. Rural households' renewable energy usage intention in Iran: extending the unified theory of acceptance and use of technology. Renew Energy 2018;122:382–91. https://doi.org/10.1016/j.renene.2018.02.011.

[27] Mahmoodi-Eshkaftaki M, Ebrahimi R. Assess a new strategy and develop a new mixer to improve anaerobic microbial activities and clean biogas production. J Clean Prod 2019;206:797–807. https://doi.org/10.1016/j.jclepro.2018.09.024.

[28] Oloko-Oba MI, Taiwo AE, Ajala SO, Solomon BO, Betiku E. Performance evaluation of three different-shaped bio-digesters for biogas production and optimization by artificial neural network integrated with genetic algorithm. Sustain Energy Technol Assessments 2018;26:116–24. https://doi.org/10.1016/j.seta.2017.10.006.

[29] Hernandez M, Rodriguez M. Hydrogen production by anaerobic digestion of pig manure: effect of operating conditions. Renew Energy 2013;53:187–92. https://doi.org/10.1016/j.renene.2012.11.024.

[30] Brindhadevi K, Shanmuganathan R, Pugazhendhi A, Gunasekar P, Manigandan S. Biohydrogen production using horizontal and vertical continuous stirred tank reactor – a numerical optimization. Int J Hydrog Energy 2021;46:11305–12. https://doi.org/10.1016/j.ijhydene.2020.06.155.

[31] Ri PC, Kim JS, Kim TR, Pang CH, Mun HG, Pak GC, et al. Effect of hydraulic retention time on the hydrogen production in a horizontal and vertical continuous stirred-tank reactor. Int J Hydrog Energy 2019;44:17742–9. doi: 10.1016/j.ijhydene.2019.05.136.